\documentclass[twocolumn,showpacs,preprintnumbers,amsmath,amssymb]{revtex4}
\usepackage[latin9]{inputenc}
\usepackage{amsmath}
\usepackage{amssymb}
\usepackage{graphicx}
\usepackage{color}
\usepackage{slashed}
\usepackage{bm}
\usepackage{epstopdf}
\usepackage[unicode=true,
 bookmarks=true,bookmarksnumbered=false,bookmarksopen=false,
 breaklinks=false,pdfborder={0 0 1},backref=false,colorlinks=false]
 {hyperref}

\makeatletter

\@ifundefined{textcolor}{}
{%
 \definecolor{BLACK}{gray}{0}
 \definecolor{WHITE}{gray}{1}
 \definecolor{RED}{rgb}{1,0,0}
 \definecolor{GREEN}{rgb}{0,1,0}
 \definecolor{BLUE}{rgb}{0,0,1}
 \definecolor{CYAN}{cmyk}{1,0,0,0}
 \definecolor{MAGENTA}{cmyk}{0,1,0,0}
 \definecolor{YELLOW}{cmyk}{0,0,1,0}

}

\makeatother

\begin{document}

\title{Higher order derivative operators as quantum corrections}

\author{L. H. C. Borges}
\email{luizhenriqueunifei@yahoo.com.br}
\affiliation{UNESP - Campus de Guaratinguet\'a - DFQ, Avenida Dr. Ariberto Pereira
da Cunha 333, CEP 12516-410, Guaratinguet\'a, SP, Brazil}

\author{F. A. Barone}
\email{fbarone@unifei.edu.br}
\affiliation{IFQ - Universidade Federal de Itajub\'a, Av. BPS 1303, Pinheirinho,
Caixa Postal 50, 37500-903, Itajub\'a, MG, Brazil}

\author{C. A. M. de Melo}
\email{cassius@unifal-mg.edu.br}
\affiliation{Instituto de Ci\^{e}ncia e Tecnologia, Universidade Federal
de Alfenas. Rod. Jos\'{e} Aur\'{e}lio Vilela (BR 267), Km 533,
n${{}^{\circ}} $11999, CEP 37701-970, Po\c{c}os de Caldas, MG,
Brazil}

\author{F. E. Barone}
\email{}
\affiliation{Brazil}

\begin{abstract}
In this paper we propose a non-minimal, and ghost free, coupling between the gauge field and the fermionic one from which we obtain, perturbatively, terms with higher order derivatives as quantum corrections to the photon effective action in the low energy regime. We calculate the one-loop effective action of the photon field and show that, in addition to the Euler-Heisenberg terms, the well known Lee-Wick term, $\sim F_{\mu\nu}\partial_{\alpha}\partial^{\alpha}F^{\mu\nu}$, arises in low energy regime as a quantum correction from the model. We also obtain the electron self energy in leading order. 
\end{abstract}

\maketitle

\section{\label{I}Introduction}

In recent years, field theories with higher order derivatives have been intensely investigated in the literature. One of the reasons to study this subject is to improve renormalization properties and to tame ultraviolet divergences. The most common theory of this kind is the so called Lee-Wick or Podolsky electrodynamics \cite{LW1,LW2,LW3,LW4,LW5,LW6} which is Lorentz and gauge invariant, unitary and also a divergence-free theory. One of the most remarkable features of this electrodynamics is the fact that it leads to a finite self energy for a point-like charge in $3+1$ dimensions \cite{LW7,LW8,LW9}. Besides, the Lee-Wick Standard Model (LWSM) has been proposed in \cite{LW10} as a theory that solves the hierarchy problem. After this work, investigations concerning it has been carried out in several different scenarios \cite{LW11,LW12,LW13,LW14,LW15,LW16,LW17,LW18,LW19,LW20,LW21,LW22}. 

Other interesting results can be found in the literature concerning Lee-Wick theories, among them we can mention, for instance, the quantization of Lee-Wick electrodynamics \cite{LW23,LW24,LW25,LWW26}, the wave propagation \cite{LW26}, the interactions between external sources \cite{LW27}, the presence of a single conducting surface \cite{LW28}, the Casimir effect in Lee-Wick electrodynamics \cite{LW29}, the non-Abelian Lee-Wick gauge theory \cite{LW30,LW31,LW32}, Lee-Wick-type theories for gravity \cite{LW33,LW34,LW34a,LW37,LW38,LW39}, supersymmetry in Lee-Wick theory \cite{LW40,LW41} and so on.

A quite natural question one can make about field theories with higher order derivatives concerns on the possible mechanisms able to generate the  Lee-Wick operators as loop quantum corrections to the photon effective action, in low energy regime. Certainly, it is be possible to construct a variety of mechanisms of this type, but we shall search for the one which deviates from QED minimally, keeping the gauge symmetry, being quadratic in the fields, and with second derivatives for the gauge field and first derivatives for the fermionic one. In this sense, we could interpret the Lee-Wick electrodynamics as an effective theory of a gauge theory which could, in principle, be free of the typical massive ghosts modes which are present in the standard Lee-Wick electrodynamics. Hence, this kind of model would exhibit some features of the Lee-Wick eletrodynamics, in low energy regime, whereas it it would be ghost free.

It is important to mention that, once we have an abelian gauge theory, the ghost field which comes from the gauge fixing shall always be present and not coupled to the other fields, so this kind of ghost is not properly a problem, on the contrary to the ghost modes inherent to the Lee-Wick theory.  

In this paper, thinking about such a mechanism, we propose a high energy gauge invariant model containing the photon field and a single massive charged fermion field. The fermion field couples to the photon one through a non-minimal term which contains derivatives of second order in the gauge field. Since we assume the fermion mass to be large, we integrate out the fermionic field and, adopting the version of the proper-time method \cite{Schwinger} (for a review, see for example\,\cite{BSD}), we obtain the one-loop low energy effective action of the gauge field. The proper-time method is a very efficient tool for obtaining the  effective action, which was adapted for gauge theories in \cite{mcarthur} and Lorentz violating theories in \cite{ourptime}. In particular, with the obtained quantum corrections, we take into account the usual vacuum polarization of the Maxwell theory, the standard nonlinear corrections to the photon, known as the Euler-Heisenberg terms in QED, and contributions of operators with higher order derivatives and quadratic in the gauge field.

We show that, just in lowest order (one loop correction), the standard Lee-Wick term arises as the quantum correction, quadratic in the gauge field in the proposed model, in low energy regime. So, we have a model which in low energy regime, recovers the same features as the Lee-Wick theory. 

This paper is structured as follows: in section \ref{II} we describe the model we compute the one-loop low-energy effective action for the photon field. We obtain the electron self energy in leading order in section \ref{CQ}. Section \ref{IV} is dedicated to our final remarks and conclusions.

Along the paper we shall deal with a 3+1 dimensional space-time and use Minkowski coordinates with the diagonal metric with signature (+, -, -, -).

\section{\label{II}The model and the effective action}

We start by proposing an {\it ad-hoc} Lagrangian containing a gauge field $A_{\mu}$ and a single massive charged fermion field $\psi$,
\begin{equation}
\mathcal{L}=-\ \frac{1}{4}F^{\mu\nu}F_{\mu\nu}+\bar{\psi}\left[i\slashed{\partial}-M-\gamma^{\mu}g\left(A_{\mu}-\frac{1}{m_{p}^{2}}\partial^{\nu}F_{\mu\nu}\right)\right]\psi\,,\label{model}
\end{equation}
where $F_{\mu\nu}=\partial_{\mu}A_{\nu}-\partial_{\nu}A_{\mu}$ is
the field strength, $M$ is the fermion mass, $g$ is the electric
charge and $m_{p}$ is a mass parameter that is responsible for a coupling between  
the fermion field with the photon field in high-energy regime. The proposed model
can be understood as a gauge-invariant theory, similar to the QED, and containing an additional non-minimal
coupling proportional to $\frac{g}{m_{p}^{2}}$.

Notice that it is the simplest modification in QED which is quadratic in the fields, keeps the gauge invariance, contains derivatives up to second order for the gauge field and first order for the fermionic one. In fact, it is the unique vector extension of the gauge connection that is gauge invariant and obtained from the gauge potential up to second order derivatives. Therefore, besides being non-minimal, this new coupling is still gauge covariant.

As we have an abelian gauge theory, when we fix the gauge, a ghost field is introduced, but it is completely decoupled from the other fields. In addition, since the fermion mass is large, in the low energy
regime we are allowed to integrate in the fermion field to obtain the one-loop correction to the effective action for the vector one.

Expressing the one-loop correction to the
effective action, $S_{eff}^{(1)}\left[A\right]$, in terms of a functional, we have 
\begin{eqnarray}
S_{eff}^{(1)}\left[A\right]&=-\,{\rm Tr}\,\ln\left(i\widetilde{\slashed D}-M\right)=\frac{1}{2}{\rm Tr}\,\ln\left(\widetilde{\slashed D}^{2}+M^{2}\right)\cr\cr
&=-\frac{1}{2}\,\mathrm{Tr\,}\ln\left(\widetilde{D}^{2}+i\Sigma^{\mu\nu}\widetilde{F}_{\mu\nu}+M^{2}\right)
\end{eqnarray}
where we defined
\begin{equation}
\widetilde{A}_{\mu}=gA_{\mu}-\frac{g}{m_{p}^{2}}\partial^{\nu}F_{\mu\nu}\ ,\label{coupling}
\end{equation}
the operator $\widetilde{D}_{\mu}=\partial_{\mu}+i\,\widetilde{A}_{\mu}$ and the matrices
\begin{equation}
\Sigma^{\mu\nu}=\frac{1}{4}\left[\gamma^{\mu},\gamma^{\nu}\right]\:;\:\widetilde{F}_{\mu\nu}=\partial_{\mu}\widetilde{A}_{\nu}-\partial_{\nu}\widetilde{A}_{\mu}\,.\label{trace2}
\end{equation}

As usual in the calculations of effective actions, we consider field configurations in low energy regime in such a way to take $F^{\mu\nu}$, aproximately, as not varying rapidly in the space and time. In this case the three operators in the argument of the lorarithm of expression (\ref{trace2}) comute each other.
 
We can compute the explicit form of the one-loop contribution to the effective action of Eq.\,\eqref{trace2} using standard methos employed in the literature to obtain the Euler-Heisenberg lagrangian, as the Schwinger method \cite{Schwinger,BaroneFarinaBoschi,BaroneAragao} or the Zete function method \cite{zeta,FarinaBaroneZeta,mcarthur,HDO}. The complete result fot the full effective action $S_{eff}\left[A\right]$ is 
\begin{eqnarray}
\label{zxc3}
S_{eff}\left[A\right]=\int d^{4}x\Biggl[-\frac{1}{4}F^{\mu\nu}F_{\mu\nu}\cr\cr
-\frac{g^{2}C}{4}F^{\mu\nu}F_{\mu\nu}
-\frac{g^{2}C}{m_{p}^{2}}F_{\mu\nu}\partial^{\alpha}\partial_{\alpha}F^{\mu\nu}\cr\cr
+\frac{g^{2}C}{2}\frac{1}{m_{p}^{4}}\partial_{\mu}\partial^{\beta}F_{\nu\beta}\partial^{\mu}\partial_{\alpha}F^{\nu\alpha}+{\cal L}_{F^{4}}\cdots\Biggr]\ .
\end{eqnarray}
where we defined the divergent constant
\begin{equation}
\label{defC}
C=\frac{1}{12\pi^{2}}\int_{0}^{\infty}d\tau\ \tau^{-1}e^{-M^{2}\tau}\ ,
\end{equation}
and ${\cal L}_{F^{4}}$ is a contribution containing terms in fourth order in the field tensor. 

Now we must renormalize the physical constants and the fields. After renormalization, we will show that the fourth term on the right hand side of Eq. (\ref{zxc3}) vanishes and the third one, is finite.

The second term in the right hand side of Eq. (\ref{zxc3}) is the one obtained in the standard QED \cite{Schwinger}. It is the vacuum polarization correction to the Maxwell theory and accounts for the field and charge renormalizations, by defining the renormalized charge and field strength, as follows
\begin{eqnarray}
g_{(ren)}^{2}&=&\frac{g^{2}}{1-g^{2}C}\cr\cr
F^{\mu\nu}_{(ren)}F^{\alpha\beta}_{(ren)}&=&\left[1-g^{2}C\right]F^{\mu\nu}F^{\alpha\beta}
\end{eqnarray}
in such a way that $g^{2}F^{\mu\nu}F^{\alpha\beta}=g^{2}_{(ren)}F^{\mu\nu}_{(ren)}F^{\alpha\beta}_{(ren)}$.

The third and fourth terms on the right hand side of Eq. (\ref{zxc3}) involve higher order derivatives of the photon field as well as the parameter $m_{p}$. We must also renormalize the Podolsky mass in such a way to avoid divergent contributions from these terms. The only way to accomplish this task is to define the renormalized Podolsky mass
\begin{equation}
\label{defmpren}
m_{p(ren)}^{2}=m_{p}^{2}/C\to\mbox{finite}\ ,
\end{equation}
as a finite quantity.

Definition (\ref{defmpren}) warrants that the third term in Eq. (\ref{zxc3}) is finite and also makes vanishing the fourth term in Eq. (\ref{zxc3}), because
\begin{equation}
\frac{C}{m_{p}^{4}}=\frac{1}{C}\frac{1}{(m_{p}^{2}/C)^{2}}=\frac{1}{C}\frac{1}{m_{p(ren)}^{4}}\to 0\ \ ,
\end{equation}
once $m_{p(ren)}$ is finite and $C$ is divergent. 

From now on, the sub-index $(ren)$ shall be implicit along the text.

In terms of normalized quantities, the action (\ref{zxc3}) reads
\begin{eqnarray}
\label{zxc4}
S_{eff}\left[A\right]=\int d^{4}x\Biggl[-\frac{1}{4}F^{\mu\nu}F_{\mu\nu}
-\frac{g^{2}}{m_{p}^{2}}F_{\mu\nu}\partial^{\alpha}\partial_{\alpha}F^{\mu\nu}\cr\cr
+{\cal L}_{F^{4}}\cdots\Biggr]\ .
\end{eqnarray}

The second term on the right hand side of Eq. \eqref{zxc4} involve higher order derivatives of the photon field and is generated from the quantum corrections of the model \eqref{model}. It is the so called Lee-Wick (or Podolsky) term, which exhibits interesting features in physics. In order to rewrite the action (\ref{zxc4}) in a more convenient form, we must also define the quantity
\begin{eqnarray}
\label{defMp}
M_{p}^{2}=m_{p}^{2}/(4g^{2})\ ,
\end{eqnarray}
so, up to second order in the gauge field, the action (\ref{zxc4}) reads
\begin{eqnarray}
\label{resultadofinal}
S_{eff}\left[A\right]=\int d^{4}x\Biggl[-\frac{1}{4}F^{\mu\nu}F_{\mu\nu}
-\frac{1}{4M_{p}^{2}}F_{\mu\nu}\partial^{\alpha}\partial_{\alpha}F^{\mu\nu}\cr\cr
-J_{\mu}A^{\mu}
+{\cal L}_{F^{4}}\cdots\Biggr]\ .
\end{eqnarray}
where we inserted the term which accounts for the interaction between the vector field and an external field source, $J^{\mu}$, which describes charges distributions \cite{BaroneHidalgo,BaroneHidalgo2,BBC,BB,BorgesBarone,BBF}. 

We point out that the result (\ref{resultadofinal}) is a one loop correction valid just in low energy regime.

Just for completeness, we point out that the fifth term on the right hand side of (\ref{zxc3}) stands for the lowest order non-linear corrections to the 
Maxwell theory, which in the case of QED is known in literature as Euler-Heisenberg term \cite{Dunne,Schwinger,TeseFabricio}. For the model (\ref{model}) we have
\begin{eqnarray}
\label{fff4}
{\cal L}_{F^{4}}=\frac{g^4}{192\pi^{2} M^{4}}\Biggl[\frac{1}{6}\left[\mathrm{Tr}\left(\widetilde{F}^{2}\right)\right]^{2}+\frac{1}{4}\left(^{*}\widetilde{F}^{\mu\nu}\widetilde{F}_{\mu\nu}\right)^{2}\cr\cr
-\frac{1}{15}\mathrm{Tr}\left(\widetilde{F}^{4}\right)\Biggr]\thinspace.
\end{eqnarray}

In the limit of large parameter $m_{p}\to\infty$, the Lagrangian \eqref{fff4} becomes the standard one obtained from QED,  namely,\begin{equation}
{\cal L}_{F^{4}}\Big|_{m_{p}\rightarrow\infty}=\frac{1}{8\pi}\frac{g^{4}}{45\pi M^{4}}\left[\left({\bf E}^{2}-{\bf B}^{2}\right)^{2}+7\left({\bf E}\cdot{\bf B}\right)^{2}\right]\thinspace.\label{eq:LEH}
\end{equation}
If the charge $g$ and the mass $M$ are taken as the electron charge $e$ and electron mass $m$ respectively,
the expression above becomes the well known Euler-Heisenberg Lagrangian in the standard QED \cite{Schwinger,TeseFabricio,f3wp}.

\section{\label{CQ} Quantum Corrections to the electron self energy}

In this section we calculate the corrections to the electron self energy in the leading order in the coupling constant $g$ and in the inverse of Podolsky mass $1/M_{p}$.

The topological structure of the Feymann diagrams of the model (\ref{model}) are the same as the ones we have in standard QED. The corresponding algebraic terms are changed according to 
\begin{eqnarray}
\label{correspondencia}
\gamma^{\mu}A_{\mu}=\gamma^{\mu}\eta_{\mu\nu}A^{\nu}
\to \gamma^{\mu} \Bigg[\eta_{\mu\nu}\Bigg(1+\frac{\Box}{m_{p}^{2}}\Bigg)-\frac{\partial_{\mu}\partial_{\nu}}{m_{p}^{2}}\Bigg] A_{\nu}
\end{eqnarray}

If we use the Lorentz gauge, we can show that the last term on the right hand side of (\ref{correspondencia}) does not contribute to the quantum amplitudes. So, we shall discard this term from now on.

For the model (\ref{model}), one can show that the electron propagator in momentum space is given by
\begin{eqnarray}
iS(p)=iS_{0}(p)+iS_{0}(p)\Big(-i\Sigma(p)\Big)iS_{0}(p)
\end{eqnarray}
where we defined the free electron propagator in momentum space 
\begin{equation}
iS_{0}(p)=i\frac{\gamma^{\mu}p_{\mu}+m}{p^{2}-m^{2}}
\end{equation}
and the electron self energy function
\begin{eqnarray}
\label{defSigma}
\Sigma(p)=-ig^2\mu^{4-d}\cr
\int\frac{d^{d}k}{(2\pi)^{d}}\biggl(1-\frac{k^{2}}{m_{p}^{2}}\biggr)\gamma_{\mu}\Biggl(\frac{{\slashed p}-{\slashed k}+M}{[(p-k)^2-M^2]k^2}\Biggr)\gamma^{\mu}
\end{eqnarray}
with the implicit limit $d\to4$.

The function (\ref{defSigma}) is composed by the standard electron self energy function, which is well known in the literature \cite{Kaku,Das}, 
\begin{eqnarray}
\label{defSigmastand}
\Sigma_{stand}(\slashed{p})=\lim_{d=4}\Bigg(\frac{g^2}{8\pi^{2}(4-d)}(-{\slashed p}+4M)\Bigg)\cr\cr
+\frac{g^{2}}{16\pi^{2}}\Bigg[{\slashed p}(1+\gamma)-2M(1+2\gamma)\cr\cr
+2\int_{0}^{1}dx[{\slashed p}(1-x)-2M]\ln\Bigg(\frac{M^{2}x-p^{2}x(1-x)}{4\pi\mu^{2}}\Bigg)\Bigg]
\end{eqnarray}
added by an additional term
\begin{eqnarray}
\label{defDelta}
\Delta(p)=ig^2\mu^{4-d}\frac{1}{m_{p}^{2}}
\int\frac{d^{d}k}{(2\pi)^{d}}\gamma_{\mu}\Biggl(\frac{{\slashed p}-{\slashed k}+M}{[(p-k)^2-M^2]}\Biggr)\gamma^{\mu}.
\end{eqnarray}
In the above expression $\gamma$ stands for the Euler constant.

With the change in the integration variable $q=k-p$, using some gamma matrices algebra, the definition (\ref{defMp}) and the fact that \cite{Das,Kaku}
\begin{eqnarray}
\int\frac{d^{d}q}{(2\pi)^{d}}\frac{1}{(q^{2}-m^{2})^{\alpha}}=\frac{i\pi(-1)^{\alpha}\Gamma(\alpha-d/2)}{(2\pi)^{d}\Gamma(\alpha)(m^{2})^{\alpha-d/2}}\ ,\cr\cr
\int d^{d}q\frac{q^{\mu}}{(q^{2}-m^{2})^{\alpha}}=0\ ,
\end{eqnarray}
we can calculate the contribution (\ref{defDelta}) by dimensional regularization
\begin{eqnarray}
\label{rfv1}
\Delta=-\lim_{d=4}\Biggl(\frac{M^{3}}{8\pi^{2}M_{P}^{2}}\frac{1}{(4-d)}\Biggr)\cr\cr
+\frac{M^{3}}{16\pi^{2}M_{P}^{2}}\Bigg[\gamma-\frac{1}{2}+\ln\Bigg(\frac{M^{2}}{4\pi\mu^{2}}\Bigg)\Bigg]\ .
\end{eqnarray}

Notice that (\ref{rfv1}) does not depend on the momentum $p$.

Collecting terms and making $d=4-\epsilon$, we have the electron self energy fuction
\begin{eqnarray} 
\Sigma(\slashed{p})=\Sigma_{stand}(\slashed{p})+\Delta\cr\cr
=\lim_{\epsilon=0}\Bigg(\frac{g^2}{8\pi^{2}\epsilon}(-{\slashed p}+4M)-\frac{M^{3}}{8\pi^{2}M_{P}^{2}}\frac{1}{\epsilon}\Bigg)\cr\cr
+\frac{g^{2}}{16\pi^{2}}\Bigg[{\slashed p}(1+\gamma)-2M(1+2\gamma)\cr\cr
+2\int_{0}^{1}dx[{\slashed p}(1-x)-2M]\ln\Bigg(\frac{M^{2}x-p^{2}x(1-x)}{4\pi\mu^{2}}\Bigg)\Bigg]\cr\cr
+\frac{M^{3}}{16\pi^{2}M_{P}^{2}}\Bigg[\gamma-\frac{1}{2}+\ln\Bigg(\frac{M^{2}}{4\pi\mu^{2}}\Bigg)\Bigg]
\end{eqnarray}

It would be interesting to investigate all the Feynman diagrams of the theory (in leading order in $g$ and $1/M_{p}$), considering the renormalizability of the theory and some vacuum effects, in a similar way that was done in Ref's. \cite{JHEP2017,PRD78CO,MedeirosBarone}.

\section{\label{IV}Conclusions and Final Remarks}

In this paper, we proposed a high energy mechanism which allowed us to obtain a term with derivatives of higher order in the gauge field in low energy regime. More specifically, we started with a high-energy model containing the photon field and a single massive charged fermion field non-minimally coupled. The one loop effective action for the gauge field was obtained integrating out the fermionic degrees of freedom with the zeta function method. In this scenario, we showed that the Lee-Wick operator emerges as a one loop quantum correction in low energy regime from the proposed mechanism, which is composed by the Dirac field coupled to the Maxwell one in a non-minimal, ghost-free and gauge-invariant way. The proposed model is the simplest deviation of QED which yields the one loop Lagrangian (\ref{resultadofinal}).

On the other hand, for a given $m_{p}$ fixed, the result (\ref{resultadofinal}) implies that the Euler-Heisenberg term (which comes from ${\cal L}_{F^{4}}$) scales as $\frac{1}{M^4}$ while Lee-Wick correction scales as $\ln(M)$ (it scales as $C\to\lim_{\mu\to\infty}\int_{1/\mu^{2}}^{\infty}d\tau e^{-M^{2}\tau}\tau^{-1}\sim\ln(M)$). Therefore, maybe it is possible (if the non-minimal coupling proposed here would be present) to exist a low energy regime such that the Bopp-Podolsky-Lee-Wick correction is manifested and more relevant in comparison with the Euler-Heisenberg one. Maybe this possibility should be considered in the analysis of experiments with high intensity lasers, such as the PVLAS experiment \cite{Lasers}.

We have also obtained the electron self energy in the leading order in the coupling constant $g$ and in the inverse of Podolsky mass $1/M_{P}$.

As a last comment, maybe the model (\ref{model}) could come out as a mechanism for which the point-like charge is finite and avoiding some possible problems exhibited by theories with higher order derivatives of the gauge field, as an indefinite metric and ghosts modes, feature found in the standard Lee-Wick electrodynamics \cite{NPB84BG}. This conjecture deserves more investigations, taking into account the perturbative study of the model (\ref{model}) in a wide analysis of the Feynman graphics of the theory, in a similar way that was done in Ref's. \cite{JHEP2017,AgliettiAnselmi,Anselmi,PRD78CO,AnselmiPiva}, and the vacuum polarization effects, as in Ref. \cite{MedeirosBarone}, for the standard QED.

\textbf{Acknowledgments.}  L.H.C. Borges thanks to S\~ao Paulo Research Foundation (FAPESP) under the grant 2016/11137-5 for financial support. F.A. Barone thanks to CNPq (Brazilian agency) under the grants 311514/2015-4 and 313978/2018-2 for financial support.

\end{document}